\documentclass{ws-procs975x65}

\begin{document}

\title{Holographic Quantum Foam}

\author{Y. Jack Ng}
\address{Institute of Field Physics, Department of Physics and Astronomy,\\
University of North Carolina, Chapel Hill, NC 27599-3255, USA\\
E-mail: yjng@physics.unc.edu}

\begin{abstract}

Due to quantum fluctuations, probed at small scales, spacetime is very
complicated --- something akin in complexity to a turbulent froth which the late
John Wheeler dubbed quantum foam, aka spacetime foam.  Our recent work
suggests that (1) we may be close to being able to detect quantum
foam with extragalactic sources once the Very Large Telescope Interferometers
(VLTI) are fully operational; (2) dark energy is arguably a
cosmological manifestation of quantum foam, the constituents of which obey 
infinite statistics; (3) in the gravitational context, 
turbulence is closely related to holographic quantum foam, partly 
validating Wheeler's picture of a turbulent spacetime.

\end{abstract}


\bodymatter

\section{Introduction}

At microscopic scales our world is known to obey quantum mechanics which is 
characterized by an
indeterminism giving rise to fluctuations in measurements.
If spacetime, like all matter and energy, undergoes quantum fluctuations, there will be 
an intrinsic limitation to
the accuracy with which one can measure a distance $l$, for that distance
fluctuates by $\delta l$.
On fairly general grounds, we expect $\delta l \gtrsim l^{1 - \alpha} l_P^{\alpha}$, \cite{ng08}
where $l_P$ is the Planck length,
the characteristic length scale in quantum gravity.
The parameter $\alpha
\sim 1$ specifies the different quantum foam models. (The canonical model \cite{mis73}
corresponds to $\alpha = 1$ with $\delta l \sim l_P$.)

Applying quantum mechanics and black hole physics (from general relativity), we 
\cite{ng94} find that a distance $l$
fluctuates by an amount $\sim l^{1/3} l_P^{2/3}$.  The corresponding quantum foam model
with $\alpha = 2/3$ has become known as the holographic model since it can be shown
\cite{ng08} to be consistent with the holographic principle \cite{tho93}.

\section{Turbulence and Holography}

There are deep similarities between the
problem of quantum gravity and turbulence \cite{jej08}.
The connection between these seemingly disparate fields is
provided by the role of diffeomorphism symmetry in classical
gravity and the volume preserving diffeomorphisms of classical
fluid dynamics.  Furthermore,
in the case of irrotational fluids in three spatial dimensions,
the equation for the fluctuations of the velocity potential can be 
written in a geometric
form~\cite{unr95} of a harmonic Laplace--Beltrami equation:
$\frac{1}{\sqrt{-g}} \partial_a( \sqrt{-g} g^{ab} \partial_b \varphi) = 0 ~$.
Here, apart from a conformal factor, the effective space time metric 
has the canonical ADM form
$ds^2 = \frac{\rho_0}{c} [ c^2 dt^2 - \delta_{ij}(dx^i - v^i dt)(dx^j - 
v^j dt)]$,
where $c$ is the sound velocity and $v^i$ are the components of the fluid's velocity 
vector.
We observe that in this expression for the metric, the velocity of the fluid $v^i$ 
plays the role of the shift
vector $N^i$ which is the Lagrange multiplier for the spatial diffeomorphism constraint 
(the momentum constraint) in
the canonical Dirac/ADM treatment of Einstein gravity:
$ds^2 = N^2 dt^2 - h_{ij} (dx^i + N^i dt) 
(dx^j + N^j dt)$.  Hence in the fluid dynamics context, $N^i \rightarrow v^i$, and a
fluctuation of $v^i$ would imply a fluctuation of the shift vector.
This is possible provided the metric of spacetime fluctuates, which is a very loose,
intuitive, semi-classical definition of the quantum foam.

But which quantum foam model?  Recall
$\delta \ell \sim \ell^{1 - \alpha} \ell_P^{\alpha}$.
If one defines the velocity as
$v \sim \frac{\delta \ell}{t_c} $,
where the natural characteristic time scale is
$t_c \sim \frac{\ell_P}{c} $,
then it follows that
$v \sim c \big(\frac{\ell}{\ell_P}\big)^{1 - \alpha} $.
It is now obvious that a Kolmogorov-like scaling \cite{k41} in turbulence 
has been obtained, i.e., the velocity scales as $v \sim
\ell^{1/3}$ and the two-point function has the needed two-thirds power law
provided that $\alpha = 2/3$.  Since the 
velocities play the role of the shifts, they describe how the metric fluctuates at 
the Planck scale.  The implication is that at short distances, spacetime is a 
chaotic and stochastic fluid in a turbulent regime \cite{wh55}
with the Kolmogorov length $l$.
The energy cascades are a property of the spacetime foam.  But we emphasize that this 
interpretation of the Kolmogorov scaling in the quantum gravitational setting is
valid only for the case of holographic quantum foam (corresponding to $\alpha = 2/3$).
\footnote{Update:  Recently
we \cite{jej09} have proposed a string theory of turbulence that explains the Kolmogorov
scaling in $3+1$ dimensions and the Kraichnan and Kolmogorov scalings in $2+1$ 
dimensions.
We argue that this string theory of turbulence should be understood from the
viewpoint of the $AdS/CFT$ dictionary.  We find that
not only can string theory be useful in formulating a theory of turbulence, but the 
physics of turbulence may provide some guidance to understanding the
quantum foam phase of strong quantum gravity.}

\section{Detectability of Quantum Foam with Extragalactic Sources}

We \cite{chr06} suggested 
that spacetime foam might be uncovered by looking for unresolved cores
in the images of distant quasars. \cite{lie03}
The point is that, due to
quantum foam-induced fluctuations in the phase velocity of an incoming light
wave from a distant point source, the wave front itself develops a small
scale ``foamy'' structure.
This results in the wave vector acquiring a cumulative random fluctuation in
direction with an angular spread of the order of
$\Delta \phi / 2 \pi$, where $\Delta \phi = 2 \pi \delta l / \lambda
= 2 \pi N l^{1-\alpha} l_P^{\alpha} / \lambda$
is the fluctuation in the phase of the electromagnetic wave with wavelength
$\lambda$ after traveling a distance $l$ from the distant source.  In
effect, spacetime foam creates a ``seeing disk'' whose angular diameter is
\cite{chr06}
$\Delta \phi /(2 \pi)
\sim (l / \lambda)^{1 - \alpha} (l_P / \lambda)^{\alpha}$. \footnote{This is
partly based on the intuition that cumulative fluctuations in lengths are 
comparable in all directions.  (In particular, we have assumed the {\it same}
cumulative factors \cite{ng08} for both transverse and longitudinal directions.)  
But we should keep in mind that this intuition, 
though reasonable, could be wrong; after all, spatial 
isotropy is here ``spontaneously" broken with the detected light 
being from a particular direction.}

For a telescope or
interferometer with baseline length $D$, this
means that dispersion (on the order of $\Delta \phi /2 \pi$ in the normal
to the wave front) will be recorded as a spread in the angular size of a
distant point source, causing a reduction in the Strehl ratio, and/or the
fringe visibility when $\Delta \phi / 2 \pi \sim \lambda / D$ 
for a diffraction limited telescope. \footnote{
For example , for a quasar of 1 Gpc away, at an infrared                               
wavelength of the         
order of 2 microns, the holographic model of spacetime foam predicts a phase 
fluctuation $\Delta
\phi \sim 2 \pi \times 10^{-9}$ radians.  On the other hand, an infrared              
interferometer with $D
\sim 100$ meters (like the VLTI) has $\lambda / D \sim 5 \times 10^{-9}$.}
Thus, in principle, for arbitrarily large distances spacetime foam sets a lower limit on 
the observable angular size
of a source at a given wavelength $\lambda$.  Furthermore, the disappearance
of ``point sources'' will be strongly wavelength dependent happening first at short
wavelengths. 
Interferometer systems (like the VLTI
when it reaches its design performance) with multiple
baselines may have sufficient signal to noise to allow for the detection
of quantum foam fluctuations.

More recently we \cite{chr09} have considered
the feasibility of using various existing or proposed telescopes
(for wavelengths from hard X-rays down to radio waves) to test the different
spacetime foam models.  Figure 1 shows the prediction
made for three different models of spacetime foam corresponding to
$\alpha = 2/3, 0.6, 1/2$ respectively, for the size of observed haloes
produced by accumulated phase dispersion for a source at two redshifts,
respectively $z=4$ and $z=1$.  The labeled arrows delineate the diffraction
limited response of various telescopes.  If the arrow
representing a telescope's diffraction limited response lies below the halo
size curve for a given $\alpha$, that model may be excluded by observations.
\footnote{Update:  Recently we \cite{chr09}
have elaborated on our proposal to detect quantum foam with extragalactic 
sources and have argued that the apppropriate distance measure for 
calculating the expected angular broadening is the total line-of-sight 
comoving distance.  We then
deal with recent data and the constraints they put on spacetime
foam models.  Thus far, images of high-redshift quasars from the Hubble Ultra-Deep Field 
(UDF) provide the most stringent test of spacetime foam theories.  Indeed, we see a
slight wavelength-dependent blurring in the UDF images selected for this study.
Using existing data in the HST archive we find it is impossible to rule out the 
$\alpha=2/3$ model, but exclude all models with $\alpha<0.65$.}
We notice that the VLTI appears to be the most promising to test the holographic model.

\begin{figure}
\begin{center}
\psfig{file=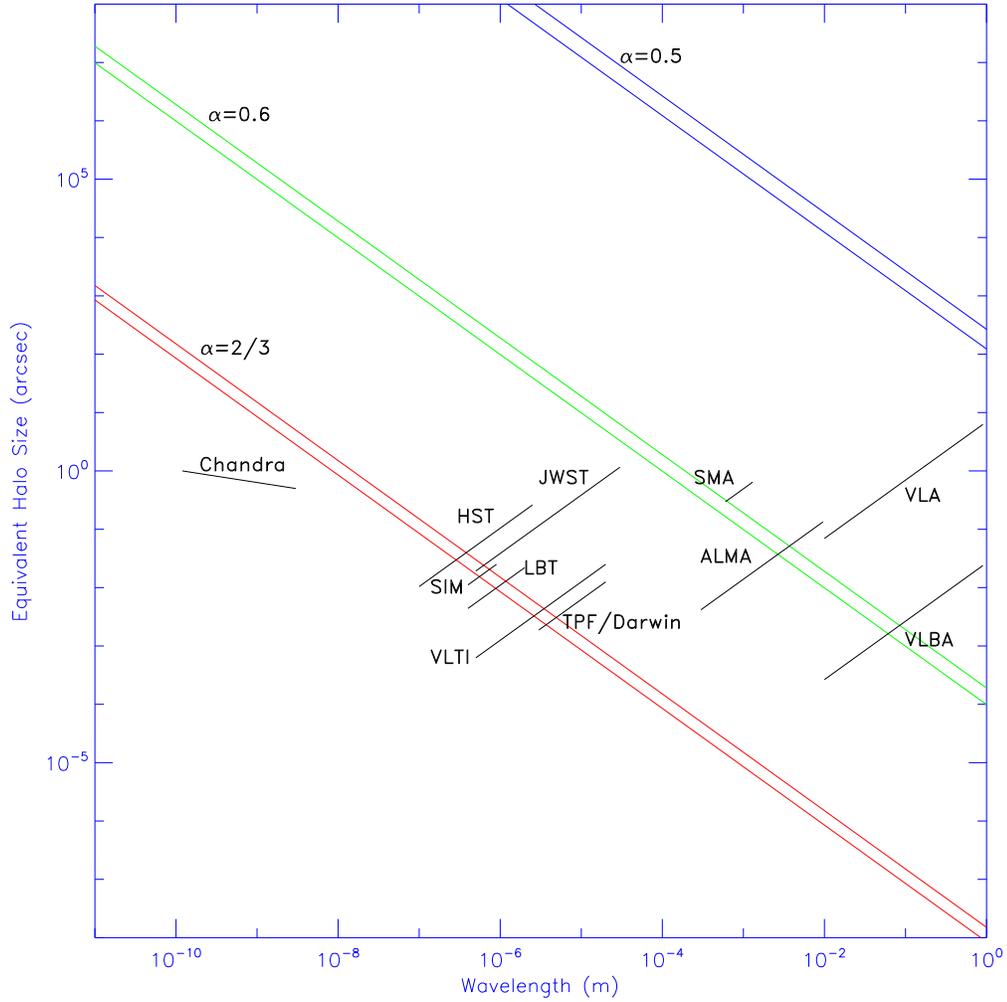,width=5.5in}
\end{center}
\caption{Detectability of various models of foamy spacetime with existing
and planned telescopes.}
\end{figure}

\section{Holographic Quantum Foam Cosmology and Infinite Statistics}

If there is unity of physics connecting the very small and the very 
large, then it is natural to apply holographic quantum foam 
physics to cosmology. \cite{ng07,ng08} 
We find that, according to the cosmololgy inspired by holographic 
quantum foam (dubbed HFC), the cosmic density is
$\rho = (3 / 8 \pi) (R_H l_P)^{-2} \sim (H/l_P)^2$, consistent with observation
($H$ is the Hubble parameter of the observable universe
and $R_H$ is the Hubble radius), \footnote{Since critical cosmic energy density
is the hallmark of the inflatonary paradigm, HFC may
supplement/implement inflation in solving some of the classic cosmological 
problems.} 
and that the cosmic entropy is given by
$I \sim (R_H /l_P)^2$.  Furthermore, with the aid of archived data from the Hubble Space
Telescope indicating the demise of the $\alpha = 1/2$ model, 
HFC has provided another argument for the existence of dark energy
 (indepenent of other cosmological/astrophysical observations of recent years). 
Successes of the conventional big bang cosmology,
such as nucleosynthesis, can also be incorporated into HFC.

Here we concentrate on one crucial question:
What is the overriding difference between conventional matter and
dark energy (perhaps also dark matter) according to HFC? 
Since dark energy carries most of the energy of the Universe, let us
focus on its constituent particles.
Consider a perfect gas of $N$ particles obeying Boltzmann
statistics
at temperature $T$ in a volume $V$.  For the
problem at hand, we take $V \sim R_H^3$, $N \sim (R_H/ l_P)^2 \gg 1$ 
and $T \sim R_H^{-1}$ (the average energy carried by each particle).
A standard calculation yields the partition function 
$Z_N = (N!)^{-1} (V / \lambda^3)^N$, where $\lambda \sim T^{-1}$.
We get, for the entropy of the system, $S = N [ln
(V / N \lambda^3) + 5/2]$.
The important point to note is that, since $V \sim \lambda^3$, the entropy
$S$ becomes nonsensically negative.
But the solution is pretty obvious: the $N$ inside the log in $S$
somehow must be absent.  In that case, the Gibbs $1/N!$ factor must be
absent from the partition function $Z_N$, accordingly the ``particles" are
distinguishable and nonidentical,
and the entropy becomes $S = N[ln (V/ \lambda^3) + 3/2]$.

Now the only known consistent statistics in greater than two space
dimensions
without the Gibbs factor is infinite statistics (sometimes called
``quantum Boltzmann statistics"). \cite{haa71,gre90}. Thus
the ``particles" constituting dark energy obey infinite statistics, instead
of the familiar Fermi or Bose
statistics. \cite{ng07}.  (Using the Matrix theory approach,
Jejjala, Kavic and Minic \cite{min07} have also argued that dark energy quanta obey
infinite statistics.)  This is the crucial difference between the constituents of
dark energy and ordinary matter, according to HFC.
Since each ``particle" has such long wavelength ($\sim R_H$), dark energy
acts like a (dynamical) cosmological constant.
But to get the correct equation of state and an appropriate transition
from an earlier decelerating to a recent accelerating cosmic expansion, 
one may need to take into account the coupling between pressureless 
dark matter and holographic dark energy. \cite{zim06}  
Finally we note that a theory of particles obeying infinite
statistics cannot be local. \cite{gre90} But intuitively holographic
theories also possess non-locality.  Thus it is not surprising that 
non-locality is present in HFC, a cosmology that is
intimately connected to holography. \cite{ng07,ng08}  The appearance of infinite 
statistics in HFC is suggestive of a holographic view of spacetime.

\section*{Acknowledgments}
This work was supported in part by the US Department of Energy.
%and the Bahnson Fund of 
%the University of North Carolina.

%\bibliographystyle{ws-procs975x65}
%\bibliography{ws-pro-sample}

\end{document}